\documentstyle[aps,preprint,floats,epsfig]{revtex}
\begin{document}
\tighten

\title{Generalized Sum Rules for Spin-Dependent \\
         Structure Functions of the Nucleon}
\author{Xiangdong Ji and Jonathan Osborne}
\bigskip

\address{
Department of Physics \\
University of Maryland \\
College Park, Maryland 20742 \\
{~}}

\date{UMD PP\#99-110 ~~~DOE/ER/40762-182~~~ May 1999}

\maketitle

\begin{abstract}
The Drell-Hearn-Gerasimov and Bjorken sum rules 
are special examples of dispersive sum rules for the
spin-dependent structure function $G_1(\nu, Q^2)$
at $Q^2=0$ and $\infty$. We generalize these sum 
rules through studying the virtual-photon Compton 
amplitudes $S_1(\nu, Q^2)$ and $S_2(\nu,Q^2)$. At small $Q^2$, 
we calculate the Compton amplitudes at leading-order
in chiral perturbation theory; the resulting sum
rules can be tested by data soon available from Jefferson Lab. 
For $Q^2>\!\!>\Lambda_{\rm QCD}^2$, the standard 
twist-expansion for the Compton amplitudes leads
to the well-known deep-inelastic sum rules. Although 
the situation is still relatively unclear 
in a small intermediate-$Q^2$ 
window, we argue that chiral perturbation theory and the 
twist-expansion alone already provide strong constraints 
on the $Q^2$-evolution of the $G_1(\nu,Q^2)$ sum rule from 
$Q^2=0$ to $Q^2=\infty$. 

\end{abstract}
\pacs{xxxxxx}

\narrowtext

\section{Introduction}

In recent years, there has been great interest 
in studying the spin-dependent structure functions
of the nucleon both in the deep-inelastic region and
in the low-energy resonance region. These interests
are spurred by a renewed effort to understand
the internal spin structure of the nucleon 
and by experimental advances in polarized beam 
and target technology. The measurements of the 
structure functions $G_1(x, Q^2)$ and $G_2(x, Q^2)$
on proton, deuteron, and He$^3$ targets in the deep-inelastic
region by the EMC, 
SMC, E142, E143, E154, E155, and HERMES collaborations
have provided rich information on the spin-dependent 
response of the nucleon for the first time \cite{data}. New 
measurements from Jefferson Lab in Virginia and 
other low-energy facilities will provide 
additional information on these interesting 
physical observables \cite{jlab}. 

A direct theoretical analysis of the 
nucleon structure functions is often difficult 
because they contain a sum over all 
nucleon final states. [An exception is the
cut-vertex formalism developed by A. Mueller in the Bjorken
region where the dominant excitations are 
few parton states \cite{mueller}.] Theoretical methods
such as chiral perturbation theory \cite{mreview} and the operator
product expansion (OPE) \cite{ope} are more suitable for 
(virtual-) photon-nucleon Compton amplitudes.  
However, for a spacelike virtual photon 
there is no known way to measure the Compton 
amplitudes directly. Fortunately, the Compton 
amplitudes and the structure functions
are related through 
dispersion relations. Hence theoretical
predictions of Compton amplitudes can be tested
against the measurable structure functions, 
provided the dispersion integrals converge.
 
The spin-dependent (virtual-) photon-nucleon
Compton process can be described by the two Compton
amplitudes $S_1(\nu,Q^2)$ and $S_2(\nu,Q^2)$, 
where $\nu$ and $Q^2$ are the virtual-photon energy 
and mass$^2$ in the target rest frame. The second 
amplitude decouples in the real photon ($Q^2=0$) limit.
Many years ago, Low, Gell-Mann, and Goldberger showed
that for a real photon the Compton 
amplitude $S_1(\nu, 0)$ reduces to 
$-\kappa^2/M^2$ in the $\nu\rightarrow 0$ limit \cite{low,gellmann},
where $\kappa$ is the (dimensionless) 
anomalous magnetic moment of the nucleon. The
result follows from electromagnetic gauge
invariance and is
nonperturbative in strong interaction physics. 
Using a dispersion relation for $S_1(\nu, Q^2)$, Drell, 
Hearn, and Gerasimov (DHG) turned this theoretical prediction into
a sum rule for the spin-dependent photonucleon
production cross sections \cite{DHG}
\begin{equation}
   \int^\infty_{\nu_{\rm in}} {d\nu\over\nu} 
(\sigma_{P}(\nu)-\sigma_{A}(\nu))  
= {2\pi^2\alpha_{\rm em}\kappa^2\over M^2},
\end{equation}
where $\sigma_{P,A}(\nu)$ is the total photonucleon 
cross section with photon energy $\nu$. The labels $P$ and $A$
refer to whether the helicity of the photon is parallel or antiparallel
to the spin of the nucleon, respectively. 
The integration begins at the inelastic threshold, denoted
here by $\nu_{\rm in}$. We note that
the spin-dependent photonucleon cross section is proportional 
to the $G_1$ structure function: $G_1(\nu, 0)
= -1/(8\pi^2\alpha_{\rm em})(\sigma_{P}(\nu)-\sigma_{A}(\nu))$.  

In the opposite limit $Q^2,\nu\rightarrow \infty$ while
$Q^2/\nu$ remains finite, Bjorken
derived a sum rule for the $G_1(\nu, Q^2)$ structure function
using the current algebra method \cite{bjorken}  
\begin{equation}
\lim_{Q^2\rightarrow\infty}\int_0^1 dx (g_1^{(p)}(x,Q^2)-g_1^{(n)}
    (x,Q^2))={g_A\over 6},
\label{bj}
\end{equation}
where $g_1(Q^2/2M\nu,Q^2) = M\nu G_1(\nu, Q^2)$ is a 
dimensionless scaling function, $g_A$ is the neutron decay constant,
and the superscripts on $g_1$ refer to the target nucleon 
(proton or neutron). The modern approach to deriving the
Bjorken sum rule starts with the operator product
expansion for the Compton amplitude $S_1(0, Q^2)$.  
It is easy to show that as $Q^2\rightarrow \infty$
\begin{equation}
    S_1^{(p)}(0, Q^2)-S_1^{(n)}(0,Q^2) = {4\over 3Q^2}g_A \ . 
\end{equation}
This result can turned into Eq. (\ref{bj})
by a simple use of the dispersion relation 
for $S_1(\nu, Q^2)$. 

The actual test of the Bjorken sum rule is not made
at $Q^2=\infty$, but at finite $Q^2$ (2 to 10 GeV$^2$)
\cite{data}.
This requires a generalization of the Bjorken 
prediction for $S_1(0, Q^2)$ to finite $Q^2$. 
The OPE is again a natural tool for this. Near $\nu=0$, 
$S_1(\nu,Q^2)$ and $S_2(\nu, Q^2)$ can be expanded in 
Taylor series in $\nu$. The coefficient 
of $\nu^n$ can be calculated as a double expansion
in $\alpha_s(Q^2)$ (QCD perturbation theory) and $1/Q^2$ 
(the twist expansion). For sufficiently
large $Q^2$ where the high-twist terms can be neglected, 
the generalization of the Bjorken sum rule involves
calculations of well-defined perturbative diagrams \cite{bjcoe}.

In this paper, we are mainly concerned with the generalization
of the Drell-Hearn-Gerasimov sum rule to non-zero $Q^2$. 
According to the above discussion, this requires a 
theoretical study of the $S_{1,2}(\nu, Q^2)$ 
amplitudes in the low $Q^2$ and $\nu$ region. We use chiral
perturbation theory because it is the natural 
tool in this kinematic regime \cite{mreview}. 
We notice that many previous works 
have used virtual-photon cross sections instead 
of the $G_1(\nu, Q^2)$
structure function to generalize the DHG integral 
\cite{bad}. We argue that these generalizations are 
neither unique nor theoretically appealing because no actual sum rule 
for the integral has been
established in these works.  

It is interesting to study the $Q^2$-evolution of 
$S_1(0, Q^2)$ as it determines the $Q^2$ evolution of
the $G_1(\nu, Q^2)$ sum rule. We will argue that 
chiral perturbation theory (the hadron description) allows us 
to understand the $Q^2$-dependence from $Q^2=0$ to the neighborhood of
0.2 GeV$^2$ and the OPE (the parton description) from $Q^2=\infty$ to the
neighborhood of 0.5 GeV$^2$. For the window between 
$Q^2=0.2$ to $0.5$ GeV$^2$, where Jefferson Lab will 
collect precise data \cite{jlab}, we do not yet have a firm 
theoretical understanding of $S_{1,2}(\nu,Q^2)$. 
Therefore, it provides an interesting ground on which to test 
various theoretical ideas.

This paper is organized as follows. In Section II, we 
specify definitions and notations. More 
importantly, we will present the dispersion 
relations and discuss their convergence.
In Section III, we consider the DHG sum rule and its 
perturbative implications, particularly in chiral
perturbation theory. In Section IV, we 
calculate $S_1(0,Q^2)$ to leading order in 
relativistic chiral perturbation theory 
and translate the result into a generalized 
DHG sum rule. We also discuss the phenomenological 
implications of the result. In section V, we 
consider a technical point--- 
the use of heavy-baryon chiral perturbation
theory in our calculations. In section VI, 
we present more general predictions 
for the $S_{1,2}(\nu,Q^2)$ amplitudes
and suggest ways to test these predictions. 
In Section VII, we focus on the $Q^2$-evolution of the 
the simplest $G_1(\nu, Q^2)$ sum rule and emphasize
the concept of parton-hadron duality.
At the same time, we point out the role 
of the nucleon elastic contribution in the sum. 
Finally, we summarize our results in Section VIII. 

\section{Dispersion Relations and Convergence}

This section contains no new results. Rather, it introduces
the various notations and definitions which will be useful in subsequent 
sections. Moreover, we state the main theoretical
assumption which relates the (virtual-) photon
Compton amplitudes to the nucleon structure functions: the 
dispersion relations. Although these relations are
not as sacred as Lorentz symmetry or 
locality, they are derived based on the fairly 
general grounds. Therefore, the attitude we take 
in this paper is that they can be used to test theoretical
predictions of the Compton amplitudes. However, 
to satisfy the purists, we will quote some standard 
arguments justifying the use of
unsubtracted integrals.

In inclusive lepton-nucleon scattering through 
one-photon exchange, the cross section depends on
the tensor \cite{ioffe}
\begin{equation}
   W^{\mu\nu} = {1\over 4\pi}
      \int d^4 \xi e^{iq\cdot\xi}\langle PS|
    [J^\mu(\xi), J^{\nu}(0)]|PS\rangle,
\end{equation}
where $|PS\rangle$ is the covariantly-normalized
ground state of a nucleon
with momentum $P^\mu$ and spin polarization $S^\mu$
and $J_\mu = \sum_i e_i \bar \psi_i\gamma_\mu \psi_i$
is the electromagnetic current (with $\psi_i$ the quark field 
of flavor $i$ and $e_i$ its charge in units of the proton charge).
The four-vector $q^\mu$ is the virtual-photon 
four-momentum.
Using Lorentz symmetry, time-reversal, and parity invariance, 
one can express the spin-dependent ($\mu\nu$ antisymmetric)
part of $W$ as
\begin{equation}
    W^{[\mu\nu]}(P, q, S) = -i\epsilon^{\mu\nu\alpha\beta} 
      q_\alpha \left[S_\beta G_1(\nu, Q^2)
     + (M\nu\, S_\beta-S\cdot q\,P_\beta)G_2(\nu, Q^2)\right] \ , 
\end{equation}
where $\epsilon^{0123}=+ 1$ and $M$ is the nucleon mass. 
$G_{1,2}(\nu, Q^2)$ are the two spin-dependent 
structure functions of the nucleon, which  
depend on two independent Lorentz scalar variables 
$Q^2=-q^2$ and $\nu=P\cdot q/M$. In the rest frame
of the nucleon, $\nu$ is the photon energy.  
[Since $S^\mu$ is present only in 
the external nucleon states, it only appears in 
the Lorentz structure, not in the scalar structure 
functions.] At times, we use the Bjorken variable $x=Q^2/2M\nu$ to 
replace the variable $\nu$. For our purposes, we assume 
$G_{1,2}(\nu,Q^2)$ are measurable in the whole 
kinematic region $0<Q^2, \nu<\infty$. 

A quantity closely-related to the hadron tensor 
is the forward virtual-photon Compton tensor, 
\begin{equation}
     T^{\mu\nu} = i\int d^4\xi e^{iq\cdot \xi}
     \langle PS|{\rm T} J^{\mu}(\xi) J^{\nu}(0)|PS\rangle \ , 
\end{equation}
which differs from $W^{\mu\nu}$ by the time ordering. 
Because of this, standard
Feynman perturbation theory can be used to calculate 
$T^{\mu\nu}$ directly. Its antisymmetric part 
can also be expressed in terms of two scalar functions:
\begin{equation}
     T^{[\mu\nu]}(P,q,S)  = -i\epsilon^{\mu\nu\alpha\beta}
      q_\alpha \left[S_\beta S_1(\nu,Q^2)
      +
    \left(M\nu\, S_\beta -S\cdot q \,P_\beta\right) S_2(\nu,Q^2)\right] \ .
\end{equation} 
We call $S_{1,2}(\nu, Q^2)$ the spin-dependent Compton 
amplitudes. 

The Compton amplitudes can be analytically 
continued to the complex-$\nu$ plane. They are real
on the real axis near $\nu=0$. They have poles
at $\nu=\pm Q^2/2M$, corresponding to $s$- and 
$u$-channel elastic scattering, and two cuts
on the real axis entending from $\nu=\pm \nu_{\rm in}$ 
to $\nu=\pm\infty$. The $s$- and $u$-channel symmetry
leads to the following crossing relations
\begin{eqnarray}
      S_1(\nu, Q^2) &=& S_1(-\nu, Q^2) \ ,
    \nonumber \\ 
      S_2(\nu, Q^2) &=& -S_2(-\nu, Q^2) \ . 
\end{eqnarray}
The $G_{1,2}(\nu, Q^2)$ structure functions are 
proportional to the imaginary parts of 
the Compton amplitudes on the real axis:
\begin{equation}
      {\rm Im}~ S_i(\nu, Q^2) = 2\pi G_i(\nu, Q^2) \ , 
\end{equation}
where $\nu$ is approaching the positive real axis
from above and the negative real axis from below.

Applying Cauchy's theorem and assuming $S_i(\nu, Q^2)$
vanishes sufficiently fast as $\nu\rightarrow\infty$, 
we can write down the dispersion relations \cite{ioffe}
\begin{eqnarray}
  S_1(\nu, Q^2) &=& 4 
   \int^\infty_{Q^2/2M}{d\nu'\nu'G_1(\nu', Q^2)
     \over \nu'^2-\nu^2} \  , \nonumber \\
  S_2(\nu, Q^2) &=& 4 \int^\infty_{Q^2/2M}{d\nu'\nu 
      G_2(\nu', Q^2)
    \over \nu'^2-\nu^2} \ , 
\label{masterdis}
\end{eqnarray}
where we have used the crossing symmetry. Using
these relations, theoretical predictions of the 
Compton amplitudes can be translated into dispersive
sum rules for the structure functions.

One question often raised about the dispersion relations 
is their convergence at high-energy. On quite general 
grounds, one would argue that the above integrals converge. 
Considering the real photon case, for instance, 
it is reasonable to expect the photoproduction
cross sections to be spin-independent at high energy. 
 
The standard practice is to resort to Regge phenomenology, 
which has actually never been tested thoroughly in 
spin-dependent and/or virtual photon scattering cases. 
According to the folklore, $G_1(\nu, Q^2)$ in the isovector channel 
behaves as $\nu^{\alpha_{a_1}(0)-1}$ at large $\nu$, 
where $\alpha_{a_1}(0)$ is the Regge intercept of 
the $a_1$ trajectory at $t=0$ (forward scattering).
Since $\alpha_{a_1}(0)$ is believed to be somewhere
between $-0.5$ and $0$, the convergence of
the integral is guaranteed. However, in the 
isoscalar channel the picture is not so clear. 
Most of the studies so far favor a convergent 
integral \cite{bass}. When the integrals do 
converge, there is still the insidious possibility 
that they do not converge to the Compton 
amplitudes \cite{ioffe}. Such a possibility is quite 
remote in QCD given the accuracy
of the experimental tests of the Bjorken sum rule \cite{data}. 
As for the $G_2(\nu,Q^2)$ integral, convergence 
apparently happens for any reasonable Regge behavior 
\cite{heimann}. 
  
Of course, independent of the theoretical guesses, 
study of the high-energy behavior of the spin-dependent 
structure functions should be 
an important part of future experimental programs. However,
in the remainder of the paper, we take it as a good working
hypothesis that the above dispersion relations are valid 
without subtraction.

\section{The Drell-Hearn-Gerasimov Sum Rule And 
Chiral Perturbation Theory}

As we have explained above, the DHG sum rule depends
on two theoretical statements : the low-energy 
theorem for the spin-dependent amplitudes \cite{low,gellmann}
and unsubtracted dispersion relation for the relevant
Compton amplitude. Although 
both are of a nonperturbative nature, it is interesting
to consider them in the context of perturbation theory.
In particular, since we are interested in the DHG sum rule
for the nucleon, we would like to consider carefully 
its significance in chiral perturbation theory. 

The low-energy theorem is related to electromagnetic
gauge invariance \cite{low,gellmann,weinberg}. It is a 
nonperturbative result which is independent
of the detailed dynamics of the bound state under 
consideration. Using the notation introduced 
above, the low-energy theorem states that for a spin-1/2
target 
\begin{equation}
     S_1(\nu,0)\rightarrow -{\kappa^2\over M^2} \  
\end{equation}
as $\nu\rightarrow 0$, 
where $\kappa$ is the anomalous magnetic moment of the 
object under consideration. 

The above result has interesting implications on the power 
series for $S_1(0,0)$ if the structure of the target can be
understood in perturbation theory. Consider, for instance, 
$S_1(0,0)$ for the electron in quantum electrodynamics. Since 
$\kappa_{\rm e} = \alpha_{\rm em}/2\pi + ...$, the leading contribution
to $S_1(0,0)$ must be of order $\alpha_{\rm em}^2$. This means
that all contributions to $S_1(0,0)$ of order $\alpha_{\rm em}$ 
must add up to zero. Indeed, if one computes $S_1(0,0)$ 
explicitly from the four diagrams shown in Fig. 1, one
finds that all contributions add up to zero.  At the next
order, one has many more diagrams. The sum of all the diagrams 
must add up to the square of Schwinger's result.

\begin{figure}
\label{fig1}
\epsfig{figure=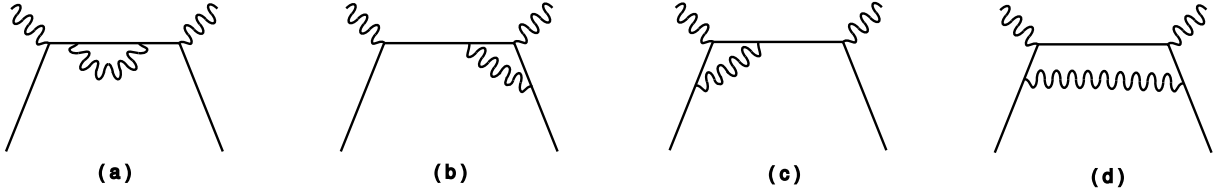,height=2.5cm}
\caption{Photon-electron Compton scattering at one-loop
order. }
\end{figure}    

\begin{figure}
\label{fig2}
\epsfig{figure=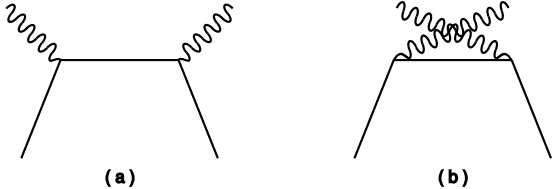,height=2.5cm}
\caption{Nucleon pole diagrams for photon-nucleon
Compton scattering.}
\end{figure}    

Now we turn to chiral perturbation theory for the low energy
structure of the nucleon. At leading order, 
the nucleon couples
to photons through the following effect vertex,
\begin{equation}
     \Gamma^\mu = e\gamma^\mu + i\kappa_0{\sigma^{\mu\nu}q_\nu\over 2M},
\end{equation}
where $\kappa_0$ is the anomalous magnetic moment of the 
nucleon in the chiral 
limit. The nucleon-pole diagrams (shown in Fig. 2) yield the following
contribution to $S_1(0,0)$ :  
\begin{equation} 
     S_1(0,0)^{N-{\rm pole}} = -{\kappa_0^2 \over M^2}.
\end{equation}
Comparing this with the low-energy 
theorem, one concludes that the leading chiral contributions to 
$S_1(0,0)$ must sum to zero. 

\begin{figure}
\label{fig3}
\epsfig{figure=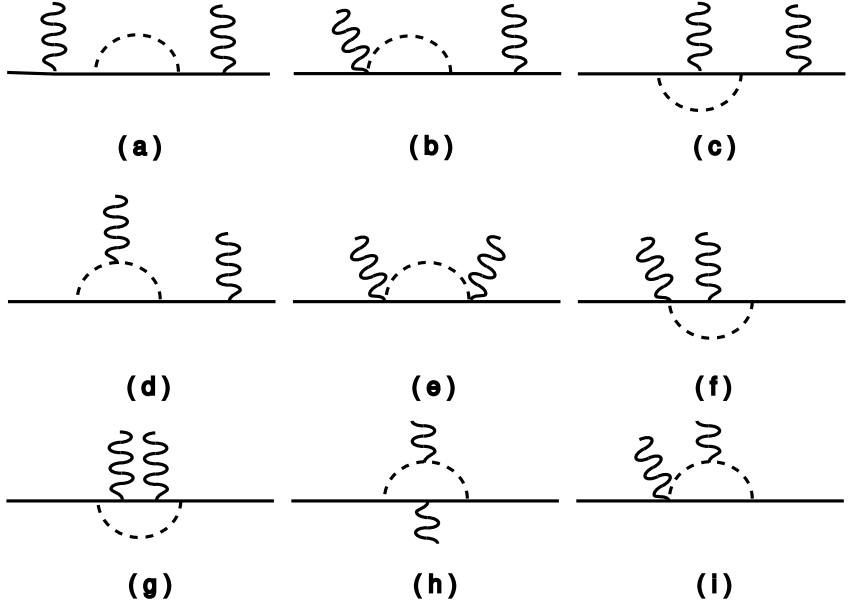,height=8cm}
\caption{Photon-nucleon Compton diagrams at one-loop 
order in chiral perturbation theory.}
\end{figure}    

We consider the one-loop contributions $S_1(0,0)$ in chiral perturbation theory
shown in Fig. 3. There are nine topologically distinct diagrams.
All others can be obtained by switching the initial and final state 
nucleon or by exchanging the photons (crossing symmetry).  All of these
diagrams contribute to $S_1(0,0)$,  as shown in Table 1.  

\begin{table}
\begin{tabular}{|c|c|c|c|c|}
Diagram & Isospin Factor & Constant & $\log(m_\pi^2/
m^2)$
 & $2/\epsilon-\log(m^2/ 4\pi\mu^2)$\\
\hline
3a & $1/2+t_3$ & $9/ 4$ & 0 & $3/ 4$\\
\hline
3b & $1/ 2+t_3$ & $-1$ & 0 & $-1$\\
\hline
3c & $1/2+t_3$ & 0 & 0 & $-1/2$\\
\hline
3d & $1/2+t_3$ & 1 & 1 & 0\\
\hline
3e & 1 & $3/ 2$ & 0 & $1/2$\\
\hline
3f & $1/ 2-t_3$ & 2 & 0 & $-1$\\
\hline
3g & $3/ 2-t_3$ & $-3/ 4$ & 0 & $1/4$\\
\hline
3h & $1/ 2-t_3$ & 1 & 1 & 0\\
\hline
3i & 1 & $-3$ & $-1$ & 0\\
\end{tabular}
\vspace{5mm}
\caption{Contributions to $S_1(0,0)$ from the diagrams in Fig. 3.  
All contributions have the common factor $2(g_A/ 4\pi f_\pi
)^2$.  We note that although the sum of each column is indeed 
zero, the cancellation is highly nontrivial.}
\label{s10}
\end{table}

We note that all diagrams without the attachment of a photon to the
pion loop are ultraviolet divergent. However, all divergences
cancel out.  The sum of all contributions
is indeed zero, as required by gauge-invariance. 

When combining the low-energy theorem with the 
unsubtracted dispersion relation, one obtains the 
DHG sum rule. If one has a fundamental theory in which
particles are pointlike, the DHG sum rule
is correct order by order, and even diagram by diagram, 
in perturbation theory. 
In QED and the standard model of electroweak interactions, 
this result has been shown at leading order in perturbation
theory \cite{altarelli}. Recently, Brodsky and Schmidt
have generalized Altarelli et al.'s result for any $2\rightarrow 2$ 
process to supersymmetric and other quantum field theories
\cite{brodsky}. 

However, it must pointed out that in an effective theory
where the particles have internal structure, the 
DHG sum rule cannot be correct order by order in 
perturbation theory. Indeed, recall the $s$-channel 
nucleon-pole diagram which contributes to $S_1(0,0)$.
This digram does not have an imaginary part because 
elastic scattering is forbidden in this case by 
kinematics (unless $\nu$ is strictly zero). However,
it gives the entire low-energy-theorem 
result in the chiral limit. The unsubtracted dispersion
relation clearly does not work for this diagram.
An opposite example is the diagram with a $\Delta$ 
pole. We will see in Sec. VI that this diagram
does not contribute to $S_1(0,0)$, but it completely
dominates the DHG integral in the limit
of a large number of color \cite{cohen}. Again, 
we see that an unsubtracted dispersion relation
fails for the diagram. In general, since the effective 
theory is only good at low-energy, it cannot be used
to calculate the photon absorption cross section
at very high-energy, and hence one cannot
establish any dispersion relation between the real
and imaginary parts of the Compton 
amplitudes in the theory. 

It is interesting to note here that a perturbative 
verification of the DHG sum rule in the context of 
quantum gravity can have implications about physics 
beyond the string compatification scale \cite{goldberg}.

\section{Generalized Drell-Hearn-Gerasimov Sum Rule}

As we have discussed above, the Drell-Hearn-Gerasimov
sum rule has its origin in the dispersion relation and the
low-energy theorem for the Compton amplitude $S_1(0,0)$. To 
extend the sum rule to virtual-photon scattering, 
we take $\nu=0$ in Eq. (\ref{masterdis})
\begin{equation}
      S_1(0, Q^2) = 4 \int^\infty_{Q^2/2M} 
     {d\nu\over\nu} G_1(\nu, Q^2) \ . 
\label{extdhg}
\end{equation}
This would be a general $Q^2$-dependent sum rule provided 
one knew how to extend the theoretical 
prediction for the Compton amplitude $S_1(0, Q^2)$ 
beyond the low-energy theorem.

We notice that the popular way to generalize
the DHG sum rule in the literature has been to 
consider an integration over the spin-dependent 
virtual-photon-nucleon cross section
(referred to as the DHG integral) \cite{bad}. 
Indeed, in some experimental publications, 
the data has been presented in this way
\cite{hermes}. [The model calculation presented
in Ref. \cite{drechsel} is an exception.]
The problem we have with this is that the 
virtual-photon cross section is not a 
well-defined quantity because the virtual-photon
flux is convention-dependent. More importantly, 
in these types of proposals, no rule has been specified 
as to what the DHG integral should be equal to. 
If one does not have a rule for the sum, there 
is no point to study the sum. One can just as 
well study the individual contributions.
The generalized DHG sum rule implied in 
Eq. (\ref{extdhg}) is not only Lorentz-invariant,
but also exhibits clear physical significance: 
The integral gives rise to the virtual-photon 
Compton amplitude $S_1(0,Q^2)$.

If $Q^2$ is very small compared with the nucleon
mass, $S_1(0,Q^2)$ should be computable in chiral 
perturbation theory. In recent years,
chiral perturbation theory ($\chi$PT) has emerged as a powerful
tool to go beyond the various low-energy theorems
involving pions, photons, and nucleons \cite{mreview,holstein}. In this 
and the next two sections, we study the 
implications of $\chi$PT for the 
$S_{1,2}(\nu, Q^2)$. We notice that a study of a generalized
DHG sum rule via a dispersion relation 
in $\chi$PT has in fact been made in 
Ref. \cite{bernard}. However, our view on the subject 
is significantly different and we 
will point out the differences towards the end of this 
section. 

In this section, we are interested in the 
leading-order chiral perturbation theory 
prediction for $S_1(0, Q^2)$.
Let us make an important point first. As soon as 
$Q^2\ne 0$, the Compton amplitude receives an 
a contribution from elastic
scattering. This contribution can be calculated 
simply in terms of the Dirac and Pauli form 
factors of the nucleon 
\begin{equation}
     S_1^{\rm el}(0,Q^2) 
   = {4\over Q^2}F_1(Q^2) (F_1(Q^2)+F_2(Q^2)) \ . 
\end{equation}
This elastic contribution dominates at low $Q^2$
due to the $1/Q^2$ singularity.
Since the DHG sum rule has no elastic
contribution due to the kinematic constraint, 
we consider a subtracted version of $S_1$
\begin{equation}
    \overline S_1(0, Q^2) = 
      S_1(0, Q^2) - S_1^{\rm el}(0, Q^2) \ , 
\end{equation}
which then can be expanded around $Q^2=0$: 
\begin{equation}  
    {\overline S_1}(0, Q^2) = - {\kappa^2\over M^2}
       + {\overline S_1}'(0,0)Q^2 + 
        {1\over 2}{\overline S_1}''(0,0) Q^4 + ... \ . 
\label{s1t}
\end{equation}
In $\chi$PT, the coefficient ${\overline S_1}'(0,0)$ 
can be expanded as a power series in pion mass: 
\begin{equation}
   {\overline S_1}'(0,0) = {a\over m_\pi^2}
     + {b\over m_\pi}(\log m_\pi + \beta) + ... \ . 
\end{equation}
The leading chiral contribution comes from the
$a$ term. 

We will calculate $a$ using
heavy-baryon chiral perturbation theory
(HB$\chi$PT) in which one formally takes the 
nucleon mass to infinity \cite{jenkins}. 
HB$\chi$PT is entirely equivalent to 
$\chi$PT using the fully relativistic
propagators except that the former has a better
way to organize different powers of the nucleon 
mass. There is, however, a subtle point
in calculating $S_1(0, Q^2)$ due to the special
kinematics. We will discuss this point
in the next section.
 
The leading chiral contribution comes from diagrams 
in which one of the two photons lands on a pion. 
This physically makes sense because in the chiral
limit, the nucleon has a pion cloud extending 
far away from it and the leading
scattering process is one in which the photon
directly interacts with the pion. 
At one-loop order, only diagrams $d, h,$ and $i$
of Fig. 3 contribute. However, when we add their
contributions together, we get
\begin{equation}
     a = 0 \ . 
\end{equation}
In other words, the leading chiral contribution
vanishes. To understand this, we recall
that in the heavy-baryon limit
spin-dependent effects vanish like $1/M$. 
This physical fact gives rise to the well-known heavy-quark
symmetry in heavy-quark physics \cite{wise}. 
In the present case, the symmetry
implies that the leading chiral terms 
$(Q^2)^{n}/m_\pi^{2n}$ vanish in Eq. (\ref{s1t}).
Our calculation confirms this. 

The above result has interesting phenomenological
implications. At present, there is no direct
data on the spin-dependent photo-production 
cross section, and therefore one cannot
make a direct test of the DGH sum rule. 
However, there is already data available 
for ${G_1}(\nu,Q^2)$ at low $Q^2$. The lowest
$Q^2$ of the data is around $m_\pi^2 \sim 0.02$ GeV$^2$. 
At this point ${\overline S_1}(0, Q^2)$ differs
from ${\overline S_1}(0,0)= -\kappa^2/M^2$ by
a quantity of order ${Q^2/m_\pi M^3}\sim
(m_\pi/M)/M^2 = 0.15/M^2$. Baring an anomalously 
large coefficient, the correction is 
of only a few percent! 

\begin{figure}
\label{fig4}
\epsfig{figure=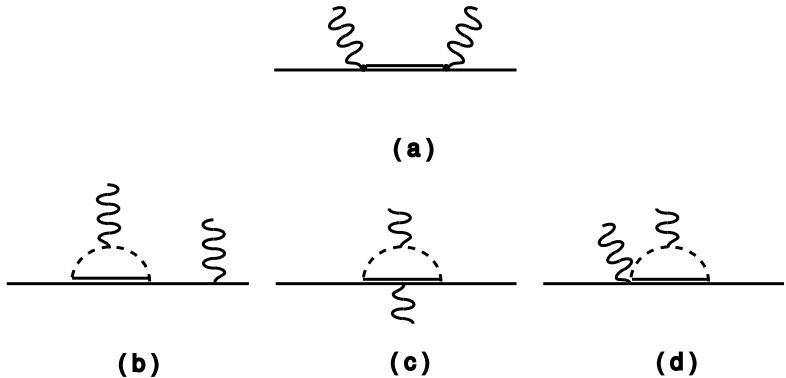,height=5cm}
\caption{Photon-nucleon Compton diagrams with
the $\Delta$ as an intermediate state.}
\end{figure}    

What about contributions from the 
$\Delta$-resonance, whose mass is about
$2m_\pi$ larger than the nucleon mass? 
In the large $N_c$ limit, the $\Delta$ is degenerate
with the nucleon and clearly plays 
a very important role. Thus, we need 
to get an idea how important the $\Delta$ is
in the real world. First of all, 
direct $\Delta$ production, shown in Fig. 4a, 
gives a contribution of zero. This is somewhat surprising
given the fact that the $\Delta$ contributes dominantly
on the dispersive side of the DHG sum rule. 
However, as we have mentioned before, one cannot
establish a dispersion relation in an effective theory 
on a diagram-by-diagram basis.  At one loop level,
we consider the diagrams shown in Figs. 4b-d.  
Our result shows that their contributions 
to $S_1(0,Q^2)$ again vanish in HB$\chi$PT, 
independent of the $\Delta$-$N$-$\pi$ coupling constant
and $\Delta$ mass. We again expect this from 
the heavy-baryon spin symmetry.

The above result strongly suggests a study 
of $S_1(0,Q^2)$ in chiral perturbation
theory at the next order. This study 
is considerably more complicated and the 
result will be presented elsewhere\cite{kao}.

A bonus of considering virtual-photon scattering
is that one has the opportunity to study another 
Compton amplitude, $S_2(0, Q^2)$. Since 
$S_2(\nu, Q^2)$ is odd in $\nu$ and also contains the 
elastic contribution, we consider a sum rule
for the subtracted ${\overline S_2}^{(1)}(0,Q^2)
= \partial(S_2(\nu,Q^2)-S_2^{\rm el}(\nu,Q^2))/\partial
\nu|_{\nu=0}$, 
\begin{equation}
         {\overline S_2}^{(1)}(0,Q^2) 
       = 4 \int^\infty_{\nu_{\rm in}}
        {d\nu\over \nu^2} G_2(\nu, Q^2) \ . 
\end{equation}
Taking into account the nucleon contribution, 
we get
\begin{equation}
   {\overline S_2}^{(1)N}(0, Q^2) = 
    4 \left({g_A\over 4\pi f_\pi}\right)^2
      \int^1_0 dx {x^2(1-2x)\over m_\pi^2+x(1-x)Q^2} \  , 
\end{equation}
at leading order in $1/M$.
One might be puzzled why ${\overline S}_2^{(1)}(0, Q^2)$
does not vanish in the heavy-nucleon limit as 
in the case of ${\overline S}_1(0, Q^2)$. The answer is that
the natural energy scale for the heavy-nucleon 
system is $1/M$. Indeed if we take $\nu\sim 1/M$,
${\overline S}_2(\nu, Q^2)$ does vanish as $M\rightarrow 
\infty$. However, when we take its derivative
with respect to $\nu$ at $\nu=0$, it is equivalent 
to taking $\nu$ of order one in ${\overline S_2}
(\nu,Q^2)$. Hence ${\overline S}_2^{(1)}(0,Q^2)$ 
survives the heavy-nucleon limit.

The $\Delta$ contribution can also be evaluated,
\begin{eqnarray}
   {\overline S}_2^{(1)\Delta}(0, Q^2) 
   &=& 2\left({4\over  3}\right)^2 \left(g_{\pi N\Delta}\over 
    4\pi f_\pi\right)^2 \int^1_0 dx 
{x^2(1-2x)\over \Delta^2-\tilde m_\pi^2}\nonumber\\
      && \times\left[1-{1\over 2}\Delta{\partial\over \partial\Delta}
{\rm Arch}^2\left(
{\Delta\over\tilde m_\pi}\right)\right],
\end{eqnarray}
where $\tilde m_\pi^2 =m_\pi^2 + x(1-x)Q^2$ and 
$\Delta$ is the mass difference between 
the $\Delta$ and the nucleon.
In the large-$N_c$ limit ($\Delta\rightarrow 0$), the 
$\Delta$ contribution
cancels exactly the nucleon's, producing a null 
result. As shown in Ref. \cite{gervais,manohar}, this type of 
cancellation must happen if the large-$N_c$ limit
produces a consistent theory. In our case, 
${\overline S}_2^{(1)}(0, Q^2)$ from nucleon intermediate
states alone diverges like $N_c$. The $\Delta$-cancellation 
ensures that the unitary is obeyed in the large-$N_c$ limit. 

In the real world of $N_c=3$, the $\Delta$
contribution to ${\overline S}_2^{(1)}(0,Q^2)$ is 
less than 20\% near $Q^2=0$. This means
that as far as ${\overline S}_2^{(1)}(0,Q^2)$ 
is concerned, the large-$N_c$ limit does not provide
a good zeroth-order approximation.

A generalization of the DHG sum rule in chiral
perturbation theory was first
investigated by Bernard, Kaiser, and Meissner \cite{bernard}. 
The main point of that work is indeed very close to 
ours, i.e., to get a generalization of the DHG sum rule through
the study of the Compton amplitudes.
However, the details here differ from those in Ref. 
\cite{bernard} in several important ways. First, 
we consider an integral which comes naturally out
of a dispersion relation for the $S_1(\nu, Q^2)$
amplitude. In Ref. \cite{bernard}, the integral
is over the virtual-photon cross section with
a special choice of flux factor. One in principle 
can consider a similar integral but with 
a different definition of flux factor. 
Second, in the large-$Q^2$
limit, the integral in Ref. \cite{bernard} 
does not yield the Bjorken sum rule.
In our case, the $G_1(\nu, Q^2)$ integral does. 
Finally, Ref. \cite{bernard}'s 
extension of the DHG integral  
gives rise to the combination $S_1(0,Q^2) - 
Q^2S_2^{(1)}(0, Q^2)$, which has a strong 
$Q^2$-dependence. In leading-order $\chi$PT, 
this $Q^2$-dependence comes entirely from 
$S_2^{(1)}(0,Q^2)$. Because of this, both sides
of the generalized sum rule no longer vanish 
in the heavy-nucleon limit---a nice physical
property of the DHG sum rule that we would like 
to keep at finite $Q^2$. 
          
\section{The Use of Heavy-Baryon Chiral \\
 Perturbation Theory At ${\bf\nu=0}$}

This section contains a discussion about a 
technical point in the use of HB$\chi$PT. We 
will show that the formalism can be used in 
the special kinematic point $\nu=0$ if one sums over 
all singular contributions. For a 
general reader, this section can be skipped. 

The kinematics associated with
the Compton amplitudes $S_{1,2}(0, Q^2)$
are special because the virtual photon 
has no energy in the rest frame of the
nucleon.  Thus it appears that one cannot 
really use HB$\chi$PT in this kinematic region. 
Consider, for instance, the nucleon-pole diagram, 
shown in Fig. 5a, in the fully relativistic 
formalism. It contains a contribution to the Compton amplitude
proportional to 
\begin{equation}
      \gamma^\mu {(\not\!P+\not\!q+M)\over 
        (P+q)^2-M^2} \gamma^\nu  \ , 
\end{equation}
where the denominator of the nucleon propagator is
\begin{equation}
    (P+q)^2-M^2 = 2P\cdot q +q^2 = -Q^2 \ . 
\end{equation}
We have used 
the kinematic condition that the nucleon 
is at rest and $q^0\equiv\nu=0$. 
Thus the pole contribution goes like $1/Q^2$,
which clearly cannot be produced through any 
local vertex in heavy-baryon 
chiral perturbation theory. 

The same problem exists for any one-particle
reducible (1PR) diagram with a single nucleon 
propagator in between the two electromagnetic
vertices: The dominant contribution 
contains a similar $1/Q^2$ piece which
seems to be beyond HB$\chi$PT. For one-particle 
irreducible diagrams, on the other hand,
no such singularity exists and we have 
checked that the fully-relativistic expression
reduces smoothly, in the heavy-nucleon limit, 
to Feynman rules in HB$\chi$PT.   
Therefore, in what follows, we focus on 
the 1PR diagrams and discuss how to 
treat them consistently in the heavy-nucleon limit. 

\begin{figure}
\label{fig5}
\epsfig{figure=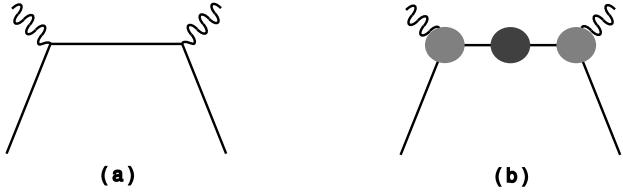,height=2.5cm}
\vspace{5mm}
\caption{Photon-nucleon Compton diagrams in fully
relativistic formalism: a) the nucleon-pole 
diagram; b) a one-particle reducible diagram. }
\end{figure}    
        
Obviously, conventional HB$\chi$PT
fails at this special kinematic point 
because the nucleon propagator in a 
1PR diagram, $1/\nu$, is divergent. 
When this happens, one has to sum over all 1PR 
diagrams to get a ``nonperturbative'' contribution.
In what follows, we are going to argue that all 
such divergent contributions just add up to 
the nucleon elastic contribution which 
we know how to calculate independently of
HB$\chi$PT. 

Let us consider a generic $s$-channel
1PR diagram shown in Fig. 5b in the usual 
fully-relativistic chiral perturbation theory. 
The structure of the diagram
is 
\begin{equation}
     \Gamma^\mu(\nu, Q^2) S_N(P+q) 
    \tilde \Gamma^\nu(\nu, Q^2) \ , 
\end{equation}
where the $\Gamma$'s represent the three-point 
nucleon-photon vertex functions and $S_N(P+q)$
is the dressed nucleon propagator. To isolate
the singular contribution, let us expand 
the $\Gamma$'s and the nucleon propagator 
at $\nu = Q^2/2M$, i.e., the onshell kinematic 
point for the intermediate 
nucleon propagator. All terms are regular 
as $\nu$ and $Q^2 \rightarrow 0$, except for
the following
\begin{equation}
      \left.\Gamma^\mu(\nu,Q^2) {i\over \not\!P+\not\!q -M} 
      \tilde\Gamma^\nu(\nu,Q^2)\right|_{\nu=Q^2/2M} \ .
\end{equation}
It is quite clear that $\Gamma(\nu=Q^2/2M,Q^2)$'s 
are just the matrix elements of the electromagnetic
currents between the on-shell nucleon states. 
Therefore all singular contributions are from the 
nucleon elastic scattering.

When all such singular contributions from 
$s$- and $u$-channel are summed, we find, for instance,
\begin{equation}
    S_1^{\rm el}(\nu, Q^2) = -2F_1(Q^2)\left(F_1(Q^2)+F_2(Q^2)\right)
    \left[{1\over 2M\nu-Q^2}
    - {1\over 2M\nu+Q^2}\right] \ , 
\label{sel}
\end{equation}
where $F_1(Q^2)$ and $F_2(Q^2)$ are the nucleon form 
factors in chiral perturbation theory. When setting 
$\nu=0$, we have a perfectly finite result. However, 
when expanded in $1/\nu$ as one does in HB$\chi$PT,
it produces $(1/\nu)^n (n=1,2,...)$ singularities. 
 
Therefore, there is no need to calculate the
$1/\nu$-type singular contributions in 
HB$\chi$PT. To account for them, one just 
calculates the Dirac and Pauli form factors 
and substitutes them in Eq. (\ref{sel}). In the
end the recipe for $S_{1,2}(0, Q^2)$ 
becomes very simple: Consider a Feynman diagram 
and calculate it at finite $\nu$ using the rules 
in HB$\chi$PT. Expand the result in a Laurent 
series about $\nu=0$, and keep only
the $\nu$-independent constant term.
Add the nucleon elastic contribution.

One final point about the HB$\chi$PT formalism. 
In the fully-relativistic calculation
presented in Sec. III, all Feynman 
diagrams contribute to $S_1(0,0)$ at order $M^0$, 
and the sum is zero because of 
electromagnetic gauge invariance. It turns
out, however, none of these diagrams 
contributes at this order in HB$\chi$PT. 
When examing the relativistic calculation carefully,
we find that the order-$M^0$ contribution comes from
the high-momentum region $k\sim M$ of the loop
integrals. In HB$\chi$PT, the leading 
contribution is nominally of order-$1/M$ 
multiplied by linearly-divergent integrals. 
These integrals can produce 
order-$M^0$ contributions in a cut-off regularization. 
In dimensional regularization, however,
the linearly-divergent pieces are set 
to zero by hand. This is allowed in field theory
because different regularization schemes 
can differ by innocuous constants 
which are finally fixed by renormalization conditions.
In the present case HB$\chi$PT
does not need to reproduce, diagram by diagram, 
the constant contributions to $S(0,0)$ in a fully-
relativistic calculation. 
The only constraint is the low-energy theorem.

\section{Extended Test of Chiral Pertubation Theory}

Chiral perturbation theory extends the 
low-energy theorems and makes general predictions 
of the spin-dependent foward Compton 
amplitudes $S_{1,2}(\nu, Q^2)$ in a kinematic region
where $\nu$ and $Q^2$ are much smaller than the 
hadron mass scale. The predictions, in particular 
the $\nu$-dependence, 
can be tested against experimental data on the 
spin-dependent structure functions $G_{1,2}(\nu, Q^2)$
through the dispersion relations. Notice that $\nu$ 
in the Compton amplitudes is not the same as the 
virtual photon-energy in $G_{1,2}(\nu, Q^2)$. 
The former $\nu$ must be kept small in order to 
justify a calculation in chiral perturbation theory, 
whereas the $\nu$-dependence 
of the structures functions must in principle 
cover the full kinematic range $0<\nu<\infty$
in order to construct the dispersion integral. 

One can think of two ways to test the $\nu$ dependence. 
First, expand $S_1(\nu,Q^2)$ and $S_2(\nu, Q^2)$ as 
Taylor series around $\nu=0$, 
\begin{eqnarray}
     S_1(\nu, Q^2) &= &\sum_{n=0,2,4,...}  
           \nu^{n}S_{1}^{(n)}(Q^2) \ ,  \nonumber \\
     S_2(\nu, Q^2) & = & \sum_{n=1,3,5,...}
          \nu^{n} S_{2}^{(n)}(Q^2) \ . 
\end{eqnarray}
Using the dispersion relations in Eq. (\ref{masterdis}), 
we find that the coefficients $S_{1}^{(n)}(Q^2)$ 
and $S_{2}^{(n)}(Q^2)$ define
two infinite towers of dispersive sum rules,
\begin{eqnarray}
    S_{1}^{(n)}(Q^2) &=& 4\int^\infty_{Q^2/2M} 
            {d\nu\over \nu^{n+1}} G_1(\nu, Q^2) ~~~~(n=0,2,4,...) \ ,
        \nonumber \\ 
    S_{2}^{(n)}(Q^2) &=& 4\int^\infty_{Q^2/2M}
            {d\nu\over \nu^{n+1}} G_2(\nu, Q^2) ~~~~(n=1,3,5,...) \ .
\label{sum5}
\end{eqnarray}
These sum rules are certainly familar in the context of 
deep-inelastic scattering \cite{jaffeji}. The point we 
want to make here is that they can be extended to 
low-$Q^2$ using the predictions for 
$S_{1,2}^{(n)}(Q^2)$ from chiral perturbation 
theory. In fact, the $S_{1}^{(2)}(Q^2)$ sum rule 
extends the known spin-dependent-polarizability 
sum rule at $Q^2=0$ to arbitrary $Q^2$ \cite{spol}. 

The second way to test the $\chi$PT  
predictions is to directly compare 
the theoretical $S_{1,2}(\nu, Q^2)$ 
with the extracted Compton amplitudes 
from data on $G_{1,2}(\nu, 
Q^2)$. We will argue below that this method 
has an important
advantage over the first one. 

We have calculated the full $Q^2$- and $\nu$-dependent 
$S_{1,2}(\nu, Q^2)$ at leading-order in  
$\chi$PT. With the nucleon intermediate
states, we need to consider only three of the diagrams 
shown in Fig. 3 ($(d)$, $(h)$, and $(i)$). 
The result is
\begin{equation}
    {\overline S_1}^N(\nu,Q^2) = 8\left({g_A\over 
       4\pi f_\pi}\right)^2
       \int^1_0 dx \left({\tilde m_\pi\over \nu}\right)
     \left[ \left(1-\left({x\nu \over \tilde m_\pi}\right)^2
     \right)^{-1/2}
     \sin^{-1}\left({x\nu\over \tilde m_\pi}\right) - {x\nu
      \over \tilde m_\pi}\right] \ . 
\end{equation}
For $x\nu>\tilde m_\pi$, one must analytically continue 
the inverse sine function and the square roots.  
Giving an infinitesimal positive imaginary part to $\nu$ yields
\begin{equation}
   {\sin^{-1}x\over\sqrt{1-x^2}}\rightarrow 
      {\ln (x-\sqrt{x^2-1})+i{\pi/2}\over\sqrt{x^2-1}} \ .
\end{equation}
As expected, the Compton amplitudes develope imaginary
parts for $\nu>m_\pi$, due to the physical production of the
$\pi N$ state. The $\Delta$ contribution comes in 
in two different ways: First through the 
pole diagram in Fig. 4a 
\begin{equation}
     S_1^{\Delta,\rm dir}(\nu, Q^2)
    =  {1\over 9}\left(G_1\over M\right)^2 
      \nu\left[{1\over \nu-\Delta+i{\Gamma_\Delta/2}}
   +{1\over \nu +\Delta - i{\Gamma_\Delta/2}}\right] \ ,  
\end{equation}
where $G_1$ is the $\gamma$-$\Delta$-$N$ coupling
(don't confuse with the structure function $G_1(\nu,Q^2)$!) 
and $\Gamma_\Delta$ is the width of the $\Delta$.  
In the large-$N_c$ limit, $G_1 = (3/\sqrt{2}) \kappa_V$ 
where $\kappa_V$ is the (dimensionless)
isovector anomalous magnetic moment of the nucleon.
The $\Delta$ also contributes via the one-loop intermediate
states in Figs. 4b, 4c, and 4d. We find 
\begin{eqnarray}
    S_1^\Delta(\nu, Q^2)
     &=& 2\left({4\over 3}\right)^2 \left({g_{\pi N\Delta}\over 
       4\pi f_\pi}\right)^2 \int^1_0 dx\,\,
     \left[x+{\tilde m_\pi\over \nu}\left(1-{\Delta\tilde\nu\over 
        \tilde m_\pi^2}\right)\right.  \nonumber \\
  &&  \times\left. 
\left(1-{\tilde\nu^2\over \tilde m_\pi^2}\right)^{-1/2}
       \left(\sin^{-1}\left({\tilde\nu\over\tilde m_\pi}\right)
-{\pi\over 2}
\right)\right]
           + (\nu\rightarrow -\nu)\,\,\, , \
\end{eqnarray}
where $\tilde \nu = \Delta-x\nu$ and $\Delta = M_\Delta-M$.
This amplitude is real when $\nu$ is below  
the pion-$\Delta$ production threshold ($<\Delta + m_\pi$).
Once again, the $\Delta$-loop contribution
cancels the nucleon contribution in the large-$N_c$
limit, and the cancellation reflects the contracted $SU(4)$
spin-flavor symmetry of the large-$N_c$ QCD \cite{gervais,manohar}. In the 
real photon limit $Q^2=0$, our result reproduces
that in Ref. \cite{bernard}. In Fig. 6a, we have shown 
the real part of the ${\overline S_1}(\nu, Q^2)$ amplitude 
in a three-dimensional plot. 
Here, we have used $g_A=1.26$ and the large-$N_c$ values 
$g_{\pi N\Delta}=3/(2\sqrt{2})g_A$ and $G_1 = (3/\sqrt{2})
\kappa_V$. The $\Delta$ resonance
contribution has been included explicitly.  
For the physical $\Delta$ mass, we find the 
contribution is small execept near 
the $\Delta$ resonance.

\begin{figure}
\epsfig{figure=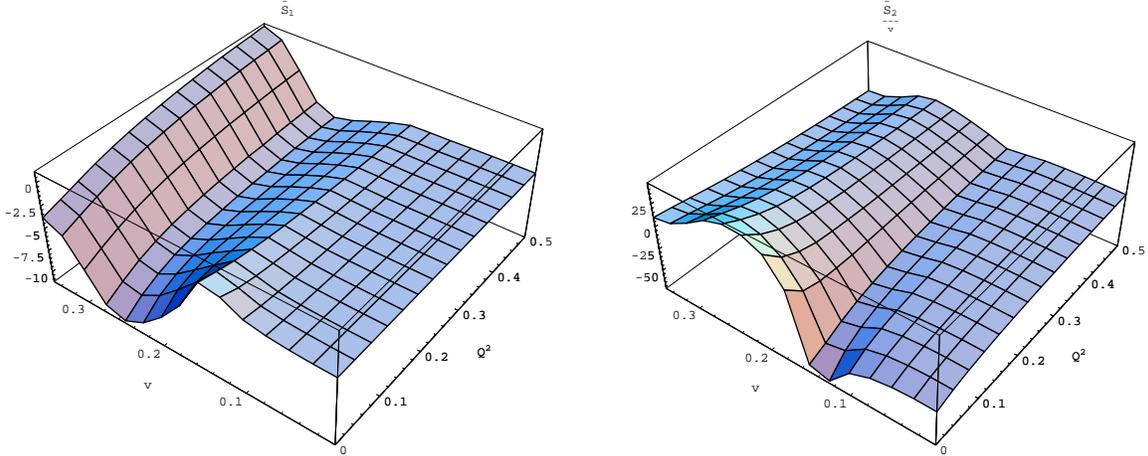, height=6.4cm}
\vspace{5mm}
\label{fig6}
\caption{(${\it Left}$)  ${\overline S_1}(\nu,Q^2)\,\, 
({\rm GeV}^{-2})$ and (${\it Right}$) 
${\overline S_2}(\nu,Q^2)/\nu\,\, ({\rm GeV}^{-4})$ 
at leading order in 
chiral perturbation theory.  $Q$ and $\nu$ are in GeV's.  We
note the N-$\pi$ and $\Delta$ resonances at $\nu=0.14\, {\rm GeV}$ 
and $0.28\, {\rm GeV}$, respectively.}
\end{figure}    

We now turn to the result for $S_2(\nu, Q^2)$. 
From the one-loop diagrams with a nucleon intermediate 
state, we find  
\begin{equation}
   {\overline S_2}^N(\nu, Q^2) = 
    4 \left({g_A\over 4\pi f_\pi}\right)^2
      \int^1_0 dx {x(1-2x)\over \tilde m_\pi}
      \left(1-\left({x\nu\over\tilde m_\pi}\right)^2\right)^{-1/2}
         \sin^{-1}\left({x\nu\over \tilde m_\pi}\right) \ , 
\end{equation}
which is odd in $\nu$. An imaginary part develops
as $\nu$ passes over the pion-production threshold.
The contribution from the direct $\Delta$ production gives
\begin{equation}
     S_2^{\Delta,\rm dir}(\nu, Q^2)
    = - {1\over 9}\left(G_1\over M\right)^2 
      \left[{1\over \nu-\Delta+i{\Gamma_\Delta/ 2}}
   +{1\over \nu +\Delta - i{\Gamma_\Delta/ 2}}\right] \ .   
\end{equation}
The one-loop diagrams with the $\Delta$ as an intermediate
state give
\begin{eqnarray}
   S_2^\Delta(\nu, Q^2) 
   &=& \left({4\over  3}\right)^2 \left(g_{\pi N\Delta}\over 
    4\pi f_\pi\right)^2 \int^1_0 dx 
{x(1-2x)\over \tilde m_\pi}
      \left(1-{\tilde \nu^2\over\tilde m_\pi^2}
      \right)^{-1/2} \nonumber \\ &&
      \times   \left[ \sin^{-1}\left({\tilde \nu\over \tilde m_\pi}\right)
     - {\pi\over 2}\right] - (\nu\rightarrow -\nu) \ . 
\end{eqnarray}
In the large-$N_c$
limit, the nucleon and $\Delta$ contribution cancel.
In Fig. 6b, we have shown a three-dimensional plot 
of ${\overline S_2}(\nu, Q^2)/\nu$. 

Because of the existence of the inelastic 
thresholds, the Taylor-expansion of 
$S_{1,2}(\nu,Q^2)$ at $\nu=0$ breaks 
down at the nearest singularity $\nu\sim m_\pi$. 
This is reflected by terms of order $(\nu/m_\pi)^n$ in 
the $\nu$-expansion. Therefore, $S_{1,2}^{(n)}(Q^2)$ are 
strongly singular and very large even for a
realistic pion mass. This is the reason that
the sum rules in Eq. (\ref{sum5}) are not the 
best way to test the $\chi$PT predictions.
Indeed, the validity of a chiral expansion
is not limited to $\nu<m_\pi$. The 
terms that we neglect in our calculation are
of order $m_\pi/M$, $\nu/M$, and $Q^2/M^2$.
The chiral expansion
is not really a Taylor expansion in $\nu$;  
the generic structure of this expansion looks like 
\begin{equation}
     S(\nu) = \sum_{n=0}^\infty {\left(\nu\over M\right)}^n 
      C_n(\nu/m_\pi) \ , 
\end{equation}
where we have neglected $Q^2$ and set $4\pi f_\pi=M$
to illustrate the point. While the coefficient
$C_n(x)$ has a zero radius of convergence when 
expanded in the chiral limit, the chiral expansion
in different orders of $\nu/M$ makes
perfect sense. 

\section{How does Drell-Hearn-Gerasimov meet Bjorken?}

Consider the nucleon structure functions $G_{1,2}(\nu, Q^2)$ 
in the deep-inelastic region: $
Q^2\rightarrow \infty$, $\nu\rightarrow \infty$, and
the ratio staying fixed. A well-known sum rule
for $g_1^{p-n}(\nu, Q^2)= M\nu G_1^{p-n}(\nu, Q^2)$
at $Q^2=\infty$ was first derived by Bjorken 
using the current algebra method \cite{bjorken}. In fact, 
Bjorken derived the more general sum rules which
are separately valid for a proton or neutron target:   
\begin{eqnarray}
   \Gamma^{p(n)}(Q^2)&\equiv&
    \int^1_0 g_1^{p(n)}(x, Q^2) dx \nonumber \\ 
   &\rightarrow &
\pm {1\over 12} g_A + {1\over 36}g_8 + {1\over 9} g_0   
\label{bj2}
\end{eqnarray}
as $Q^2\rightarrow\infty$,
where $g_A$, $g_8$, and $g_0$ are the nucleon matrix
elements of the axial currents $A^\mu_a$ $(a=3, 8, 0)$, 
\begin{equation}
    \langle PS| A^\mu_a(0) |PS\rangle = g_a (2S^\mu) . 
\end{equation}
The matrix element $g_A$ is precisely known from 
neutron-$\beta$ decay, and $g_8$ may be extracted 
from hyperon-$\beta$ decay if chiral SU(3) 
symmetry in under control\cite{manohar}. The simple quark
model predicts $g_0\equiv \Delta\Sigma=1$. 
Ellis and Jaffe assumed $\Delta\Sigma=g_8$ based on the
empirical observation that the strange quarks play an 
insignificant role in nucleon structure 
\cite{ellis}. Ellis and Jaffe's estimate for the right-hand side
of Eq. (\ref{bj2}) has been called the Ellis-Jaffe
sum rule. We note that the Bjorken sum rule can
also be derived using the modern technique of 
operator product expansions in QCD.  

To reconcile the discrepency between the EMC
extraction of $\Delta \Sigma$ at ${\overline Q^2}=10$ 
GeV$^2$ from the polarized 
deep-inelastic scattering data and the simple 
quark model prediction, Anselmino et al. observed that the 
$g_1^p(x,Q^2)$ integral at finite $Q^2$ can have 
significant deviation from the Bjorken's 
prediction \cite{anselmino}. Their argument is as follows:
Consider the $Q^2$-dependence
of the integral
\begin{eqnarray}
  I(Q^2)&\equiv& \int^\infty_{Q^2/2M} 
M^2G_1(\nu, Q^2) d\nu/\nu \nonumber \\
  &=& {2M^2\over Q^2} \Gamma(Q^2) \ . 
\end{eqnarray} 
As $Q^2\rightarrow \infty$, the Bjorken sum rule indicates 
that $I^p(Q^2)$ approaches zero from above. On the other hand, 
the DHG sum rule says $I^p(0) = -\kappa^2_p/4$, large and negative.  
Therefore, $I(Q^2)$ must undergo 
a rapid variation from $Q^2=\infty$ to 0. This 
variation could imply a significant deviation of the
the EMC data at ${\overline Q^2}=10$ GeV$^2$ from the 
Bjorken sum rule at $Q^2=\infty$. They devised a simple
phenomenological model to illustrate this. 

As pointed out by one of us \cite{ji1} and explained
in more depth below, Anselmino et al.'s observation 
is misleading in its original intent. However, 
undertanding the $Q^2$-evolution of $\Gamma(Q^2)$ or 
$I(Q^2)$ over the full range of $Q^2$ is interesting
in itself. We will argue below that
the $Q^2$-dependence can largely be understood
in terms of chiral perturbation theory
at low $Q^2$ and the operator production expansion
at high $Q^2$. There exists, however, a small 
intermediate-$Q^2$ window where the transition
from hadron to parton degrees of freedom occurs
and where we do not yet have a firm 
theoretical handle. Future precise experimental 
data in this region could help us to understand 
how the parton-hadron transition happens.

In the definition of the $I(Q^2)$ integral in 
Ref. \cite{anselmino}, the lower integration limit 
indicates that the nucleon elastic 
contribution was included. However, since 
the elastic contribution is absent at $Q^2=0$, 
overwhelming at small $Q^2$, and finally becomes 
negligible at high $Q^2$, $I(Q^2)$ cannot vary 
continuously from $Q^2=0$ to $\infty$
unless the elastic contribution is subtracted. 
Therefore, the $Q^2$-evolution picture considered 
by Anselmino et al. does {\it not} actually include 
the elastic contribution. For a quanity integrating over
inelastic contributions only, there
is no twist expansion, and hence one cannot 
say anything about the size of the twist-four
matrix element from Anselmino et. al.'s
simple phenomenological model. The same comment
applies to Ref. \cite{ib}. 
  
To really understand the $Q^2$-evolution of 
$I(Q^2)$ from $Q^2=0$ to $\infty$, we again consider 
the $Q^2$-dependent dispersive sum rule, 
\begin{equation}
        S_1(0, Q^2) = 4\int^\infty_{Q^2/2M} 
           {d\nu\over \nu} G_1(\nu, Q^2) \ ,
\label{kk} 
\end{equation}
which, as emphasized above, is the 
origin of the generalized DHG sum rule and 
the Bjorken sum rule. Consequently,$I(Q^2)$ 
or $\Gamma(Q^2)$ as a function of $Q^2$ can be obtained
just from the Compton amplitude $S_1(0, Q^2)$.
Since the available theoretical methods, 
such as chiral perturbation theory,
operator product expansion, or lattice QCD are 
natural tools for calculating $S_1(0,Q^2)$, 
it is natural to consider the $G_1$-integral 
including the elastic contribution. We know of no way 
to directly calculate ${\overline S_1}(0, Q^2)$ 
other than using the definition ${\overline S_1}(0,Q^2) 
\equiv S_1(0,Q^2) - S_1^{\rm el}(0,Q^2)$. 

Once the elastic contribution is included, we see 
that it dominates $S_1(0,Q^2)$ at low $Q^2$ because
of the $1/Q^2$ singularity. Since $S_1(0,Q^2)$
contains the same $1/Q^2$ factor at 
asymptotic $Q^2$, it is convenient to 
study just the $Q^2$ evolution of 
\begin{equation}
     \Gamma(Q^2) 
     = {Q^2\over 8}S_1(0, Q^2) \ , 
\end{equation}
It interesting to note that for a pointlike proton, 
$\Gamma(Q^2)$ is simply 1/2, independent of $Q^2$. 

In the $Q^2=\infty$ to $0.5$ GeV$^2$ region, 
$\Gamma(Q^2)$ can be obtained essentially from parton physics
formalized in the operator product expansion. From  
the OPE for the Compton amplitude $S_1(0, Q^2)$, we can write 
\begin{equation}
  \Gamma(Q^2) = \sum_{\tau=2,4,...}
     {\mu_\tau(Q^2) \over (Q^2)^{(\tau-2)/2}} \ ,
\end{equation}
where $\mu_\tau$ itself is a perturbation
series in $\alpha_s(Q^2)$. For instance \cite{bjcoe}  
\begin{eqnarray}
    \mu_2(Q^2) &=& \left(1-{\alpha_s\over \pi}
   -3.58\left({\alpha_s\over \pi}\right)^2
   -20.22\left({\alpha_s\over \pi}\right)^3\right)
    \left(\pm{g_A\over 12}+{a_8\over 36}\right) + \nonumber \\
 && + \left(1-0.33{\alpha_s\over \pi}-0.55\left({\alpha_s\over\pi}\right)^2
-4.45\left({\alpha_s\over\pi}\right)^3\right){\Delta\Sigma\over 9}
    + {\cal O}(\alpha_s^4).
\end{eqnarray}
The $\mu_4$ correction has been expressed in 
terms of various nucleon matrix elements \cite{ji2}. No one
has yet worked out the local operators associated
with $\mu_6$. The physical significance 
of the twist expansion is that at large $Q^2$, 
the Compton amplitude can be understood in terms of 
the scattering of a few partons.

What is the physical scale parameter which controls 
the twist expansion? 
By studying the structure of higher-twist operators,
one finds that the relevant dimensional parameter
is the average parton transverse momentum 
in the nucleon \cite{georgi}. On dimensional
grounds, one can write
\begin{equation}
      \mu_\tau \sim \langle k_\perp \rangle^{\tau-2} \ . 
\end{equation}
The numerical size of $\langle k_\perp\rangle$ 
is believed to be around 0.4 to 0.5 GeV. Therefore,
the naive expectation is that the $1/Q^2$-expansion
is a good approximation down to $Q^2\sim 0.5$ GeV$^2$. This
picture is in fact consistent with the 
experimental data. As $Q^2$ changes from $0.5$ to 
$\infty$, the $x$-dependence of the $g_1(x, Q^2)$ scaling function 
changes significantly. However, its first moment, $\Gamma(Q^2)$, 
is nearly independent of $Q^2$, as required by the 
twist expansion. In Ref. \cite{ji3}, 
the twist-four matrix elements were extracted from the
$Q^2$ dependence of data on $\gamma(Q^2)$. In Fig. 7a, 
we have plotted the evolution of $\Gamma(Q^2)$ for 
the proton with the twist-two and twist-four
contributions included ($\mu^p_4$ is taken to be 
0.04 GeV$^2$).  

At low $Q^2$, the elastic contribution dominates
$\Gamma(Q^2)$, which approaches $e(1+\kappa)/2$
as $Q^2\rightarrow 0$. As $Q^2$ deviates from
zero, $\Gamma(Q^2)$ falls rapidly due to the 
nucleon elastic form factor. The leading inelastic
contribution appears as a linear term in $Q^2$
and the slope can be obtained from the DHG sum rule.  
The sign of this slope enhances the rapid 
decrease of the elastic contribution.
Chiral perturbation theory allows us to 
compute the $Q^2$ dependence of the inelastic 
contribution as a perturbation series in $Q^2$
\begin{equation}
   {\overline \Gamma}(Q^2) = - {\kappa^2Q^2\over 8M^2}
   + {\cal O}\left({Q^4\over ((4\pi f_\pi)^2Mm_\pi)}\right) \ . 
\end{equation} 
The total $Q^2$ variation is 
\begin{equation}
    \Gamma(Q^2) = {1\over 2}F_1(Q^2)\left(F_1(Q^2)
    +F_2(Q^2)\right) + {\overline \Gamma}(Q^2) \ . 
\end{equation}
In Fig. 7a, we have plotted the elastic contribution 
as the dashed curve and the elastic plus the leading
term in ${\overline \Gamma}(Q^2)$ as a solid curve. Adding higher-order 
chiral corrections will bring changes to the solid 
curve. However, chiral power counting
guarantees that the change in the $Q^2$ range 
between 0 to approximately $0.2$ GeV$^2$ will be relatively 
small. Therefore, we believe that the description of $\Gamma(Q^2)$
in terms of the hadron degrees of freedom will be  
trustable at least up to $Q^2=0.2$ GeV$^2$. 

\begin{figure}
\label{fig7}
\epsfig{figure=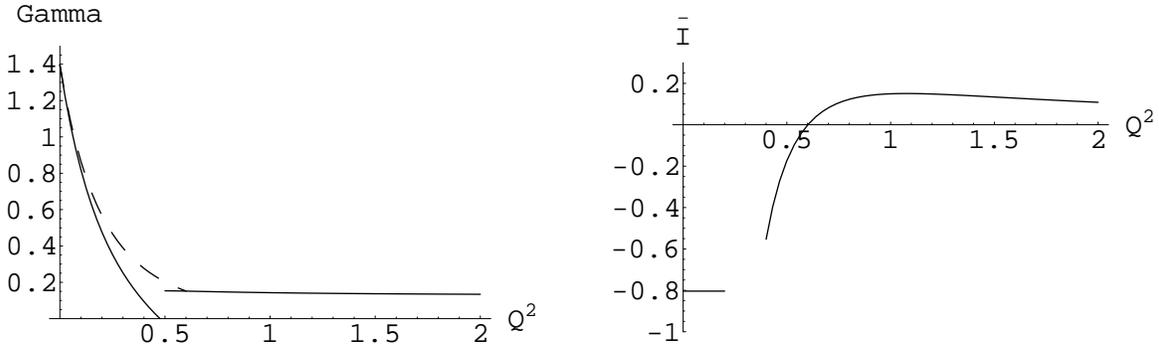, height=4.8cm}
\vspace{5mm}
\caption{${\it Left}$ : $Q^2$-evolution of the 
$\Gamma(Q^2)$ integral studied 
from the Compton amplitude $S_1(0, Q^2)$. 
${\it Right}$ : The subtracted integral ${\overline I}(Q^2)$.}     
\end{figure}    

As Fig. 7a shows, the $Q^2$-dependence of $\Gamma^p(Q^2)$ 
can already be understood over a large kinematic
region in terms of chiral perturbation theory and
the twist expansion. Mathematically, there is little
left to the imagination between $Q^2=0.2$ and 0.5 GeV$^2$
other than a smooth curve connecting the high and low-$Q^2$
regions. Physically, however, the missing link 
is very interesting. This is the regime in which 
the transition between parton and hadron degrees
of freedom happens and so it provides an important 
ground on which to test ideas about how the 
transition occurs. There is a possibility that chiral
perturbation theory and the twist-expansion 
with the appropriate inclusion of higher-order terms (e.g.
pade approximation, resonance dominance, etc. )
are simultaneously valid in an overlapping region. 
This scenario has been the main theoretical condition 
for the validity of QCD sum rule calculations 
\cite{sumrule}. Alternatively, it
is possible that there is an open window in which
neither theory works. In this case, one must
consider other methods, such as lattice QCD. 
It is interesting to note that $S_1(0, Q^2)$
can in fact be calculated in lattice QCD by rotating 
the Minkowski time in Eq. (6) to Euclidean.

The global $Q^2$-variation of $\Gamma(Q^2)$ reveals
a simple and appealing physical picture. When the photon 
momentum is low, it scatters coherently over all 
the constituents of the nucleon and the response 
shows a diffractive type of peak which stands well above 
the pointlike value of 1/2.
When $Q^2$ is large, the photon sees the individual
quarks in the nucleon and the scattering is an incoherent
sum of all charged constituents. Thus, $\Gamma(Q^2)$
is necessarily much smaller than 1/2. In fact, it
will asymptotically vanish if the proton contains 
no pointlike particles but a smeared charge 
distribution. The transition between the whole proton
and its constituent scattering happens at $Q^2$ 
corresponding roughly to the size of the system. 

The only unattractive feature of Fig. 7a is that 
the elastic contribution at low $Q^2$ overwhelms the 
inelastic contribution, making the transition
region less dramatic. Therefore, we plot in Fig. 7b 
the $Q^2$-dependence of the subtracted integral 
${\overline I}(Q^2)$, as suggested by Anselmino et al.
Now, we see the our leading chiral prediction 
becomes a straight line which we draw up to 0.2 GeV$^2$. 
On the high-$Q^2$ side, we find that ${\overline I}(Q^2)$ crosses
from positive to negative at $Q^2 \sim 0.6$ GeV$^2$. The
significance of this crossing point is that it is where the elastic
contribution equals the prediction of the leading twist expansion. 
Below this $Q^2$, the curve dives down to meet the 
low-energy theorem $(-\kappa^2/4)$ and its chiral extension. 
Therefore, while the detail may vary (the crossing point is
likely less than 0.5 GeV$^2$ according to E143 data \cite{data}), 
the seemingly intriguing $Q^2$ variation 
of ${\overline I}(Q^2)$ near the crossing point is nothing but
a simple consequence of an innocent construct.
The higher-order corrections in chiral perturbation 
theory and the twist-expansion may
provide a more accurate account of how the Drell-Hearn-Gerasimov
sum rule finally meets that of Bjorken.

\section{Summary and Outlook}

In this paper, we have studied generalized sum rules
for the spin-dependent structure functions 
of the nucleon. We emphasize that to 
obtain sum rules, one has to find ways to 
calculate the virtual-photon forward
Compton amplitudes $S_{1,2}(\nu,Q^2)$. 

Away from the $\nu=Q^2=0$ point where there is a 
well-known low-energy theorem, one can use chiral 
perturbation theory to calculate the $\nu$ and $Q^2$ 
dependence. We find that for $S_1(0,Q^2)$ the 
leading chiral contribution vanishes because of the
spin symmetry of the heavy-nucleon limit. The 
result from the next-order $\chi$PT calculation,
for example, the coefficient of the $Q^2/(Mm_\pi(4\pi f_\pi)^2)$ 
term, will be published elsewhere \cite{kao}. Our 
result for ${\overline S_2}^{(1)}(0, Q^2)$ yields a new sum rule
for the $G_2(\nu,Q^2)$ structure function. 
The complete $Q^2$ and $\nu$ dependence of the
low-energy Compton amplitudes can be tested 
through standard dispersion relations. 

For large $Q^2$, the Compton amplitudes can 
be expressed as a power series in $1/Q^2$
following an operator product expansion of the 
electromagnetic currents.
We emphasize that this is possible only for 
the complete amplitudes including the nucleon
elastic contribution. For $S_1(0, Q^2)$, 
the twist expansion is expected to converge 
for $Q^2>0.5$ GeV$^2$. Using
dispersion relations,
the Bjorken sum rule can be generalized approximately
down to this $Q^2$. 

Therefore, the low-energy $Q^2$
dependence of $S_1(0,Q^2)$ and the related sum rule
can be described in terms of the physics of the hadron 
degrees of freedom, whereas the high-energy dependence
can be described in terms of parton degrees of freedom. 
According to what we know of both theories, 
there is only a small $Q^2$ window in which 
we have no solid theoretical understanding. Clearly,
precise experimental data in this region 
will allow us to understand how the transion
from parton to hadron degrees of freedom happens. 
On the theoretical side, it is important to
pursue higher-order corrections in chiral perturbation
theory and the twist expansion. Ultimately, a precision
lattice QCD calculation may help establish
the final missing link between the DHG and 
Bjorken sum rules.

\acknowledgements
We thank J. P. Chen and T. Cohen for a number of discussions
related to experimental and theoretical aspects of the paper.  
This work is supported in part by funds provided by the
U.S.  Department of Energy (D.O.E.) under cooperative agreement
DOE-FG02-93ER-40762.

\end{document}